\documentstyle[preprint,aps,axodraw,epsf]{revtex}
\draft
\begin{document}

\title{ \vspace{3.cm}{\normalsize
   \begin{flushright}
    CERN-TH/97-288\\[-0.3cm]
    MPI/PhT/97-60\\[-0.3cm]
    hep-ph/9710380\\[-0.3cm]
    September 1997
  \end{flushright} 
}
Effective Charge of the Higgs Boson}

\author{Joannis Papavassiliou$^a$ and Apostolos Pilaftsis$^b$}

\address{$^a$Theory Division, CERN, CH-1211 Geneva 23, Switzerland\\
$^b$Max-Planck-Institut f\"ur Physik, F\"ohringer Ring 6, 80805 Munich, 
    Germany}

\maketitle

\begin{abstract} 
  The  Higgs-boson  lineshape is  studied  within the  pinch technique
  resummation formalism.  It is shown   that any resonant  Higgs-boson
  amplitude contains   a  universal part  which  is gauge independent,
  renormalization-group   invariant,  satisfies     the  optical   and
  equivalence theorems, and  constitutes the natural extension  of the
  QED effective charge to the case of the Higgs scalar.
\medskip

\end{abstract}

\newpage

\medskip
The production of the Standard Model (SM) Higgs boson \cite{Higgs} and
the detailed study  of its lineshape, mass and  width, are expected to
dominate the particle physics scene for the next two decades.  A Higgs
boson with mass $M_H$ less than 100 GeV can  be discovered at the CERN
Large Electron Positron collider LEP2 through the Bjorken process $e^+
e^- \to   ZH$.  If the  Higgs   boson turns  out   to be  heavier, its
discovery will become again possible at the CERN Large Hadron Collider
through a variety of sub-processes,  such as $X\to H^{*}\to X'$, where
$X,X'  = t\bar{t}, ZZ, W^+W^-$.  Depending  on the value  of $M_H$ and
the specific kinematic circumstances, any of the above transitions may
be resonant.  The phenomenological importance  of the above  processes
makes  the need for  solving  a subtle theoretical problem  \cite{VW},
namely the  self-consistent treatment of  the Higgs boson resonance in
the framework of S-matrix perturbation theory,  all the more pressing.
In particular, a resummation formalism needs be devised which complies
with a  set  of   very stringent   and  tightly  interlocked  physical
requirements.  To any finite  order  in perturbation theory,  physical
amplitudes reflect the  local  gauge symmetry, respect  unitarity, are
invariant under the renormalization group, and satisfy the equivalence
theorem \cite{EqTh,CG}. All of  the  above properties should be   also
present after  resummation; unfortunately,  resummation  methods often
end  up violating   one or  more of  them,  essentially because subtle
cancellations are distorted  when certain parts  of  the amplitude are
resummed  to  all  orders  in  perturbation   theory, whereas  others,
carrying important physical   information,  are only considered to   a
finite order.
 
Recently however  \cite{PP}, a formalism based  on the pinch technique
(PT) \cite{PT}   has been developed,  which   manifestly preserves the
crucial  physical  properties  during all intermediate   steps  of the
resummation procedure.   The PT  algorithm rearranges systematically a
given amplitude into {\it physically meaningful} sub-amplitudes, which
have the same kinematic properties as their conventional counterparts,
but none of  their individual pathologies.   In this Letter, the above
formalism  is extended to the  case  of resonant transitions involving
the SM Higgs boson.  The main novel results of our  study are: (i) The
PT gives rise to a Higgs boson self-energy  which is {\em independent}
of  the gauge-fixing parameter (GFP) in  every gauge fixing scheme, is
{\em universal} in the sense  that it is {\em process-independent}, it
may be  {\em  resummed}   following the  method presented   in   Ref.\ 
\cite{PP},  it displays   only  {\em physical}  fermionic and  bosonic
thresholds, and  satisfies {\em individually}  the optical theorem for
{\em both} fermionic as well  as bosonic contributions.  (ii) When the
resummed  Higgs    boson propagator is  multiplied   by  the universal
quantity $g^2_w/M^2_W$, or, equivalently, by the inverse square of the
vacuum expectation value (VEV) of the Higgs  field, it gives rise to a
{\em renormalization group invariant}  quantity, in direct analogy  to
the {\em effective charge}  of the photon,  or the $W$ and $Z$  bosons
\cite{PRW}, which constitutes a common  component in every Higgs-boson
mediated process, and can be viewed as  a physical entity intrinsic to
the Higgs   boson.  (iii)    Any amplitude   involving  longitudinally
polarized gauge  bosons satisfies   the equivalence theorem,   but its
individual $s$-channel and $t$-channel contributions do not.  Instead,
the  PT   rearrangement   of  such an amplitude    gives   rise to two
kinematically  distinct pieces, a  genuine  $s$-channel and a  genuine
$t$-channel, which satisfy the equivalence theorem {\em individually}.
In particular, the above property {\em  persists} even {\em after} the
$s$-channel Higgs boson self-energy has  been resummed, thus solving a
long-standing problem.

We  shall now  analyze the above   points  in the context of  specific
examples.    When  the    center-of-mass  (c.m.)   energy   $\sqrt{s}$
approaches $M_H$,  amplitudes  containing  an  $s$-channel Higgs boson
become singular,  and must be  regulated.  The naive  extension of the
standard Breit-Wigner procedure  to    this  case would consist     of
replacing the  free  Higgs  boson  propagator $\Delta_H  (s) =    (s -
M^2_H)^{-1}$ by   a resummed  propagator  of  the form   $[s - M^2_H +
\Pi^{HH}(s)]^{-1}$, where  $\Pi^{HH}(s)$  is the one-loop  Higgs boson
self-energy.   However,  bosonic   radiative corrections  induce    an
additional   dependence on  the GFP,  as  one  can verify  by explicit
calculations   in   a variety   of conventional gauges,    such as the
renormalizable  ($R_{\xi}$),   or  axial  gauges.   Turning   to  more
elaborate gauge fixing  schemes does not improve  the  situation.  For
example,     within  the background  field   gauges   (BFG's) and with
irrelevant  tadpole graphs   omitted,   the contribution of  the   $Z$
boson-loop reads: \cite{PPHIGGS}
\begin{eqnarray}
  \label{DBFG}
\Pi^{\widehat{H}\widehat{H}}_{(ZZ)}(s,\xi_Q)\ &=&\ \frac{\alpha_w}{32\pi}\,
\frac{s^2}{M^2_W}\, \Big\{
\Big(\, 1\, -\, 4\, \frac{M^2_Z}{s}\, +\, 12\, \frac{M^4_Z}{s^2}\Big)
B_0(s,M^2_Z,M^2_Z)\nonumber\\
&&- \Big[ 1\, +\, 4\xi_Q\, \frac{M^2_Z}{s}\, -\, (M^2_H+4\xi_QM^2_Z)\,
\frac{M^2_H}{s^2}\, \Big]\, B_0(s,\xi_Q M^2_Z,\xi_Q M^2_Z)\, \Big \}\, ,
\end{eqnarray}
where $\alpha_w=g^2_w/(4\pi)$ is the weak  fine structure constant and
$B_0$ is  the usual Passarino-Veltman  function.   The presence of the
GFP $\xi_Q$ results in bad high energy  behavior and the appearance of
unphysical  thresholds,  as   can  be verified  directly    using $\Im
mB_0(s,M^2,M^2) =  \theta(s-4M^2) \pi (1-4M^2/s)^{1/2}$.   Even though
to  any  order   in perturbation    theory  physical  amplitudes   are
GFP-independent,  and   display   only physical  thresholds, resumming
$\Pi^{\widehat{H}  \widehat{H}}_{(ZZ)}   (s,\xi_Q)$     will introduce
artifacts   to the  resonant  amplitude.   Even in   the unitary gauge
($\xi_Q \to\infty$),   where only   physical  thresholds survive,  the
$s^2$-growth in Eq.\ (\ref{DBFG})  grossly contradicts the equivalence
theorem.

\begin{figure}
   \leavevmode
 \begin{center}
   \epsfxsize=14.0cm
    \epsffile[0 0 454 340]{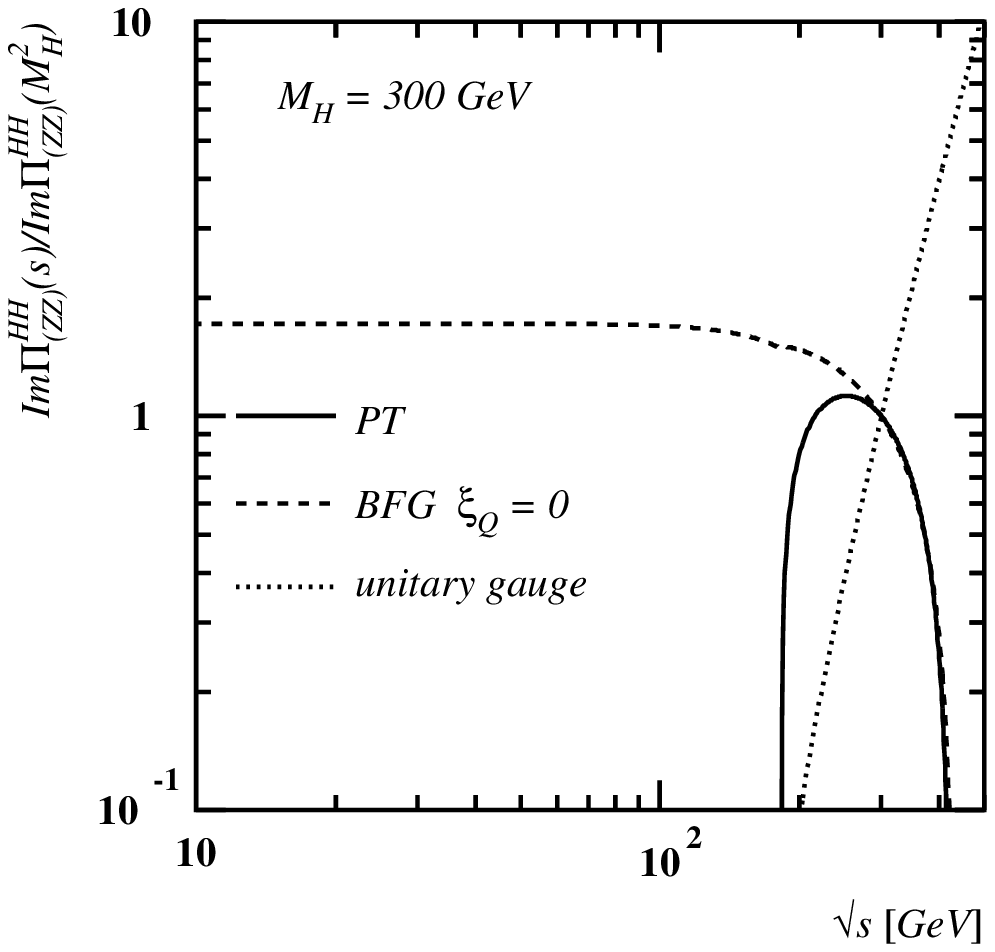}
 \end{center}
 \vspace{-0.5cm} 
\caption{ Dependence of $\Im  m\Pi^{HH}_{(ZZ)}(s)/\Im
  m  \Pi^{HH}_{(ZZ)}(M^2_H)$  on $s^{1/2}$   in  the PT,  the BFM with
  $\xi_Q = 0$, and the unitary gauge.}\label{f1}
\end{figure}

In  the PT framework however, a  modified one-loop self-energy for the
Higgs boson  can  be constructed,  by   appending to the  conventional
self-energy additional  propagator-like contributions concealed inside
vertices and      boxes.   These contributions      can  be identified
systematically, by resorting exclusively to elementary Ward identities
of  the  form $\not\!  k  (v  + a\gamma_5) = (\not\!     k + \not\!  p
-m)(v+a\gamma_5)-  (v-a\gamma_5)   (\not\!  p  -m)   + 2a  m\gamma_5$,
triggered by the longitudinal virtual momenta $k_\mu$.  Following this
procedure, we find the PT Higgs-boson self-energy \cite{PPHIGGS}
\begin{equation}
  \label{HPT}
\widehat{\Pi}^{HH}_{(ZZ)}(s)\ =\ \frac{\alpha_w}{32\pi}\frac{M_H^4}{M_W^2}
\Big[\, 1+4\frac{M_Z^2}{M_H^2}- 4\frac{M_Z^2}{M_H^4}
(2s - 3M_Z^2)\, \Big]\, B_0(s,M^2_Z,M^2_Z)\, ,
\end{equation}
which  is   GFP-independent in any   gauge   fixing scheme,  universal
\cite{NJW}, grows linearly with $s$,  and displays physical thresholds
only.  For illustration, in Fig.\ \ref{f1}, we  plot the dependence of
the running width, $\Im m\Pi^{HH}_{(ZZ)}(s)$, on $\sqrt{s}$ within the
PT resummation formalism, the  BFG with $\xi_Q  = 0$, and the  unitary
gauge.  The difference in the phenomenological predictions between the
three approaches is rather striking, in accordance with the discussion
given above.

The PT self-energies  satisfy the optical theorem {\it  individually},
as        explained   in      \cite{PP,PRW}.     To     verify    that
$\widehat{\Pi}^{HH}_{(ZZ)}(s)$    has this   property, consider    the
tree-level  transition   amplitude ${\cal   T}(ZZ)$  for the   process
$f(p_1)\bar{f}(p_2)\to  Z(k_1)Z(k_2)$; it is the  sum of an $s$- and a
$t$-  channel contribution, denoted   by ${\cal T}^H_s(ZZ)$ and ${\cal
  T}_t(ZZ)$, respectively, given by
\begin{eqnarray}
  \label{THsZZ}
{\cal T}^H_{s\, \mu\nu} (ZZ)\ &=&\ \Gamma^{HZZ}_{0\mu\nu}\, 
\Delta_H (s)\ \bar{v}(p_2) \Gamma^{Hf\bar{f}}_0 u(p_1)\, ,\\
  \label{TtZZ}
{\cal T}_{t\, \mu\nu} (ZZ)\ &=&\ \bar{v}(p_2)\Big( 
\Gamma^{Zf\bar{f}}_{0\nu}\, \frac{1}{\not\! p_1 + \not\! k_1 - m_f}\, 
\Gamma^{Zf\bar{f}}_{0\mu}\, +\, 
\Gamma^{Zf\bar{f}}_{0\mu}\, \frac{1}{\not\! p_1 + \not\! k_2 - m_f}\,
                            \Gamma^{Zf\bar{f}}_{0\nu} \Big)u(p_1)\, .
\end{eqnarray}
Here,   $s=(p_1+p_2)^{2}=(k_1+k_2)^2$  is the   c.m.\  energy squared,
$\Gamma^{HZZ}_{0\mu\nu}   =  ig_w\,     M^2_Z  /  M_W     g_{\mu\nu}$,
$\Gamma^{Hf\bar{f}}_0   =    -i    g_w\,  m_f    /   (2    M_W)$   and
$\Gamma^{Zf\bar{f}}_{0\mu}\,  =\, -ig_w/(2c_w)\, \gamma_\mu\, [  T^f_z
(1 - \gamma_5) - 2Q_fs^2_w]$, with $c_w = \sqrt{1 - s^2_w} = M_W/M_Z$,
are  the   tree-level  $HZZ$,  $Hf\bar{f}$  and $Zf\bar{f}$ couplings,
respectively, and $Q_f$ is the electric charge of the fermion $f$, and
$T^f_z$ its $z$-component of the weak isospin.   We then calculate the
expression  $[{\cal  T}^H_{s\, \mu\nu}  (ZZ) +  {\cal  T}_{t\, \mu\nu}
(ZZ)]  Q^{\mu\rho}(k_1)Q^{\nu\sigma}(k_2) [{\cal T}^H_{s\, \rho\sigma}
(ZZ)+{\cal     T}_{t\,   \rho\sigma} (ZZ)]^*$,  where  $Q^{\mu\nu}(k)=
-g^{\mu\nu}+k^{\mu}k^{\nu}/M^2_Z$    denotes  the usual   polarization
tensor, and   isolate  its Higgs-boson mediated  part.   To accomplish
this,  one must first    use  the longitudinal  momenta   coming  from
$Q^{\mu\rho}(k_1)$  and  $Q^{\nu\sigma}(k_2)$ in  order to extract the
Higgs-boson part of ${\cal T}_{t}^{\mu\nu}(ZZ)$, {\em i.e.},
\begin{equation}
  \label{TP}
\frac{k_1^\mu k_2^\nu }{M^2_Z}\, {\cal T}_{t\,\mu\nu}(ZZ)\ =\
{\cal T}_{P}^{H}+ \dots\ =\ -\, \frac{ig_w}{2M_W}\ 
\bar{v}(p_2)\Gamma^{Hf\bar{f}}_0 u(p_1)\ + \dots \, ,
\end{equation}
where   the   ellipses denote   genuine  $t$-channel  (not Higgs-boson
related)  contributions.  Then,  one    must append the   piece ${\cal
  T}_{P}^{H}$  ${\cal  T}_{P}^{H*}$ to   the ``naive'' Higgs-dependent
part    ${\cal    T}^H_{s\,    \mu\nu}   (ZZ)$      $Q^{\mu\rho}(k_1)$
$Q^{\nu\sigma}(k_2)$      ${\cal   T}^{H*}_{s\,  \rho\sigma}    (ZZ)$.
Integrating the expression so  obtained over the two-body phase space,
we finally arrive at the imaginary  part of Eq.\ (\ref{HPT}), which is
the announced result.

The   gauge-invariance of  the   S   matrix imposes  tree-level   Ward
identities  on  the unrenormalized    one-loop  PT Green's   functions
\cite{PT,JP}.  The requirement that the same Ward identities should be
maintained after renormalization leads to important QED-type relations
for the  renormalization constants  of the  theory.   Specifically, we
find
\begin{equation}
  \label{ZWZH}
\widehat{Z}_W\ =\ \widehat{Z}_{g_w}^{-2}\, ,\quad
\widehat{Z}_Z\ =\  \widehat{Z}_W \widehat{Z}_{c_w}^2\, ,\quad
\widehat{Z}_H\ =\ \widehat{Z}_W\, (1+ \delta M^2_W/M^2_W)\, ,
\end{equation}
where $\widehat{Z}_W$,   $\widehat{Z}_Z$, and $\widehat{Z}_H$  are the
wave-function renormalizations of  the   $W$,  $Z$ and  $H$    fields,
respectively, $\widehat{Z}_{g_w}$ is the coupling renormalization, and
$\widehat{Z}_{c_w}=(1+\delta    M_W^2/M_W^2)^{1/2}(1+\delta  M_Z^2   /
M_Z^2)^{-1/2}$. The  renormalization of the bare  resummed Higgs-boson
propagator $\hat{\Delta}^{H,0}(s)$ proceeds as follows:
\begin{equation}
  \label{DeltH}
\hat{\Delta}^{H,0}(s)\, =\, [\,s - (M^0_H)^2 +
\widehat{\Pi}^{HH,0}(s) ]^{-1}\, =\, \widehat{Z}_H [\,s - M_H^2 +
\widehat{\Pi}^{HH}(s)]^{-1}\, =\, \widehat{Z}_H\,
\hat{\Delta}^H(s)\, , 
\end{equation}
with $(M^0_H)^2 = M_H^2 +  \delta M_H^2$. The renormalized Higgs-boson
mass $M_H^2$   may be defined  as the  real part  of  the complex pole
position    of  $\hat{\Delta}^H(s)$.   Notice   that   within  the  PT
resummation formalism  the gauge-independent    pole \cite{AS}  of   a
resonant transition amplitude does  not get shifted \cite{PP}, and the
$HZ$   mixing is absent  up to  two  loops \cite{APRL}.  Employing the
relations in Eq.\ (\ref{ZWZH}), we observe that the universal quantity
\begin{equation}
  \label{RGIC}
  \widehat{R}^{H,0}(s)\  =\  \frac{(g^0_w)^2}{(M^0_W)^2}\,
  \hat{\Delta}^{H,0}(s) \ =\  \frac{g^2_w}{M_W^2}\, \hat{\Delta}^{H}
  (s)\ =\ \widehat{R}^H(s)
\end{equation}
is invariant   under   the  renormalization  group.  This    important
universal property of the  Higgs boson is  true for  non-Abelian gauge
theories with spontaneous symmetry  breaking (SSB), but does not  hold
in general.   For example,  in pure scalar    theories any attempt  to
construct  quantities  analogous to  $\widehat{R}^H  (s)$ fails  to be
process independent \cite{PPHIGGS}.  In that sense, $\widehat{R}^H(s)$
provides a natural extension of the notion of the QED effective charge
for the SM Higgs boson, {\em i.e.}, $H$  couples universally to matter
with an effective ``charge'' inversely proportional to its VEV. In the
high-energy  limit, $s\gg\  M^2_H$,    the dispersive  part of   Higgs
self-energy behaves as $\Re e  \widehat{\Pi}^{HH} (s) \sim -\alpha_w s
\ln   (s/M^2_H) (3m^2_t - 4M^2_W -   2M^2_Z) /(8 \pi  M^2_W)$.  If the
heavy   top   quark  were  assumed    to be  absent,  the  coefficient
accompanying the leading  logarithm in $\Re  e \widehat{\Pi}^{HH} (s)$
would be positive.  This feature is  reminiscent of the PT self-energy
in pure Yang-Mills theories  \cite{PT,PP,PRW}, whose leading logarithm
is  proportional to $b_1 =  11 c_A/3 > 0$, where  $c_A$ is the Casimir
eigenvalue  of the   adjoint   representation,  thus  reflecting   the
asymptotic  freedom of  the  theory.  On  similar theoretical grounds,
$\Im m   \widehat{\Pi}^{HH}_{(ZZ)}   (s)$ turns  negative  for   c.m.\
energies much  higher  than $M_H$, {\em  viz.},  the Higgs self-energy
cannot be spectrally represented.

\begin{figure}

\begin{center}
\begin{picture}(400,200)(0,0)
\SetWidth{0.8}

\ArrowLine(0,130)(30,100)\ArrowLine(30,100)(0,70)
\DashArrowLine(30,100)(50,100){3}\GCirc(65,100){15}{0.8}
\DashArrowLine(80,100)(100,100){3}\GCirc(115,100){15}{0.8}
\Photon(120,115)(140,130){3}{2}\Photon(120,85)(140,70){3}{2}
\Text(10,140)[r]{$f (p_1)$}\Text(10,60)[r]{$\bar{f}(p_2)$}
\Text(160,140)[r]{$Z_\mu (k_1)$}\Text(160,60)[r]{$Z_\nu (k_2)$}
\Text(40,110)[]{$H$}\Text(90,110)[]{$H$}
\Text(65,125)[]{$\hat{\Delta}^H(q)$}
\Text(135,100)[l]{$\Gamma^{HZZ}_{0\mu\nu} + \widehat{\Gamma}^{HZZ}_{\mu\nu}$}
\LongArrow(130,132)(122,124)\LongArrow(130,68)(122,76)

\Text(65,20)[]{\bf (a)}

\ArrowLine(240,120)(270,120)\ArrowLine(270,120)(270,80)
\ArrowLine(270,80)(240,80)\Photon(270,120)(300,120){2}{4}
\Photon(270,80)(300,80){2}{4}
\Text(240,130)[l]{$f$}\Text(240,70)[l]{$\bar{f}$}
\Text(275,100)[l]{$f (p_1+k_1)$}
\Text(300,130)[r]{$Z_\mu$}\Text(300,70)[r]{$Z_\nu$}

\Text(270,20)[]{\bf (b)}

\ArrowLine(340,120)(370,120)\ArrowLine(370,120)(370,80)
\ArrowLine(370,80)(340,80)\Photon(370,120)(400,120){2}{4}
\Photon(370,80)(400,80){2}{4}
\Text(340,130)[l]{$f$}\Text(340,70)[l]{$\bar{f}$}
\Text(375,100)[l]{$f (p_1+k_2)$}
\Text(400,130)[r]{$Z_\nu$}\Text(400,70)[r]{$Z_\mu$}

\Text(370,20)[]{\bf (c)}

\end{picture}
\end{center}
\caption{Resummation of the Higgs-mediated amplitude pertinent to 
$f\bar{f} \to ZZ$.}\label{f2}
\end{figure}
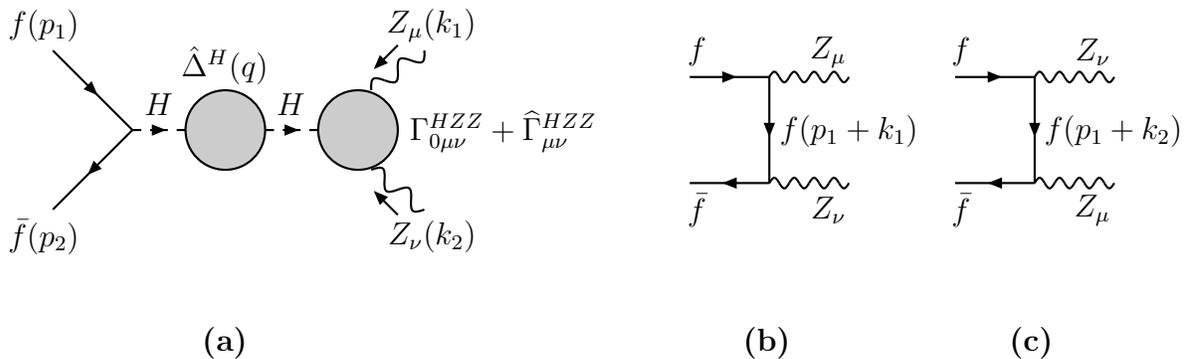

An  additional,  highly non-trivial  constraint,  must  be  imposed on
resummed amplitudes; they have  to obey the (generalized)  equivalence
theorem (GET),  which  is known to  be  satisfied  before resummation,
order by order in perturbation  theory.   For the specific example  of
the amplitude ${\cal T} (ZZ)={\cal T}^H_s+{\cal T}_t$ , the GET states
that
\begin{eqnarray}
  \label{GETZZ}
{\cal T}(Z_LZ_L)\ =\  -\, {\cal T}(G^0G^0) \,  
-i\,{\cal T}(G^0 z)\, -i \,{\cal T}(z G^0)\, +\, {\cal T}(z z)\, ,
\end{eqnarray}
where $Z_L$ is  the longitudinal component  of the $Z$ boson, $G^0$ is
its  associated    would-be    Goldstone boson,  and     $z^\mu  (k) =
\varepsilon^\mu_L(k) - k^\mu/M_W$ is the energetically suppressed part
of  the longitudinal polarization  vector  $\varepsilon^\mu_L$.  It is
crucial  to observe,  however, that already    at the tree level,  the
conventional $s$- and  $t$- channel sub-amplitudes  ${\cal T}^H_s$ and
${\cal T}_t$ fail  to satisfy the GET  individually.  To  verify that,
one   has  to   calculate ${\cal   T}_s^H  (Z_LZ_L)$,  using  explicit
expressions for the   longitudinal polarization vectors, and check  if
the  answer obtained is  equal to the Higgs-boson mediated $s$-channel
part  of the LHS  of Eq.\ (\ref{GETZZ}).   In particular, in the c.m.\
system, we have $z^\mu (k_1) = \varepsilon^\mu_L (k_1) - k_1^\mu/M_Z =
- 2M_Zk^\mu_2/s    +  {\cal  O}(M^4_Z/s^2)$,  and exactly    analogous
expressions  for $z^\mu (k_2)$.  The  residual  vector $z^\mu (k)$ has
the properties    $z_\mu k^\mu  =  -M_Z$  and  $z^2  = 0$.     After a
straightforward calculation, we obtain ${\cal T}_s^H (Z_LZ_L) = -{\cal
  T}^H_s (G^0G^0) -i {\cal T}^H_s (zG^0) -i {\cal T}^H_s(G^0z) + {\cal
  T}^H_s (zz) - {\cal T}^H_P$, where
\begin{eqnarray}
  \label{THsGG}
{\cal T}^H_s (G^0G^0)\ &=&\  \Gamma_0^{HG^0G^0}\,  
\Delta_H (s)\ \bar{v}(p_2)\Gamma^{Hf\bar{f}}_0 u(p_1)\, ,\\
  \label{THsZG}\nonumber
{\cal T}^H_s (zG^0)\, +\, {\cal T}^H_s(G^0z)\ &=&\ 
[z^\mu (k_1)\, \Gamma^{HZG^0}_{0\mu}\, +\, 
z^\nu (k_2)\, \Gamma^{HG^0Z}_{0\nu}]\, \Delta_H (s)\
\bar{v}(p_2)\Gamma^{Hf\bar{f}}_0 u(p_1)\nonumber\, ,
\end{eqnarray}
and ${\cal T}^H_s  (zz)= z^\mu (k_1)z^\nu (k_2) {\cal T}^H_{s\,\mu\nu}
(ZZ)$,   with  $\Gamma^{HG^0G^0}_0   =    -i  g_w   M^2_H/(2M_W)$  and
$\Gamma^{HZG^0}_{0   \mu}    = -  g_w  (k_1  +   2  k_2)_\mu /(2c_w)$.
Evidently,  the  presence of  the term ${\cal  T}^H_P$ prevents ${\cal
  T}_s^H (Z_LZ_L)$  from satisfying the GET.    This is not surprising
however, since an  important Higgs-boson mediated $s$-channel part has
been  omitted.    Specifically,   the    momenta  $k_{1}^{\mu}$    and
$k_{2}^{\nu}$ stemming from  the  leading  parts of the   longitudinal
polarization        vectors          $\varepsilon^\mu_L(k_1)$      and
$\varepsilon^\nu_L(k_2)$    extract    such  a   term    from   ${\cal
  T}_t(Z_LZ_L)$.   Just as happens  in Eq.\  (\ref{TP}), this term  is
precisely  ${\cal T}^H_P$,   and  must be    added  to  ${\cal  T}_s^H
(Z_LZ_L)$, in  order  to form a  well-behaved  amplitude at very  high
energies.   In other  words,  the  amplitude $\widehat{{\cal   T}}^H_s
(Z_LZ_L)   = {\cal T}^H_s (Z_LZ_L) +   {\cal T}^H_P$ satisfies the GET
independently ({\em cf.}\  Eq.\ (\ref{GETZZ})).  In fact, this crucial
property   persists after  resummation.  Indeed,   as  shown in  Fig.\
\ref{f2}, the  resummed amplitude $\overline{\cal T}^H_s (Z_LZ_L)$ may
be constructed from ${\cal  T}^H_s (Z_LZ_L)$ in Eq.\ (\ref{THsZZ}), if
$\Delta_H  (s)$ is   replaced by the  resummed Higgs-boson  propagator
$\hat{\Delta}^H  (s)$,    and $\Gamma^{HZZ}_{0  \mu\nu}$   by  the
expression $\Gamma^{HZZ}_{0\mu\nu} + \widehat{\Gamma}_{\mu\nu}^{HZZ}$,
where $\widehat{\Gamma}_{\mu\nu}^{ HZZ}$  is the one-loop $HZZ$ vertex
calculated  within the PT  \cite{PPHIGGS}.  It is then straightforward
to  show that  the   Higgs-mediated amplitude $\widetilde{\cal  T}^H_s
(Z_LZ_L) = \overline{\cal T}^H_s (Z_LZ_L)\, +\, {\cal T}^H_P$ respects
the  GET {\em individually}; to that  end we only   need to employ the
following tree-level-type PT WI's:
\begin{eqnarray}
  \label{PTHZZ1}
k^\nu_2 \widehat{\Gamma}_{\mu\nu}^{HZZ}
(q,k_1,k_2) + iM_Z \widehat{\Gamma}^{HZG^0}_\mu (q,k_1,k_2)
&=& -\, \frac{g_w}{2c_w}\, \widehat{\Pi}^{ZG^0}_\mu (k_1)\, ,\nonumber\\
  \label{PTHZZ2}
k^\mu_1 \widehat{\Gamma}_\mu^{HZG^0}
(q,k_1,k_2) + iM_Z \widehat{\Gamma}^{HG^0G^0}(q,k_1,k_2)
&=& -\, \frac{g_w}{2c_w}\, \Big[\, \widehat{\Pi}^{HH}(q^2) + 
\widehat{\Pi}^{G^0G^0}(k^2_2)\, \Big]\, ,\\
  \label{PTHZZ3}
k^\mu_1 k^\nu_2 \widehat{\Gamma}_{\mu\nu}^{HZZ}
(q,k_1,k_2) + M^2_Z \widehat{\Gamma}^{HG^0G^0}(q,k_1,k_2)
&=& \frac{ig_wM_Z}{2c_w}\, \Big[\, \widehat{\Pi}^{HH}(q^2) + 
\widehat{\Pi}^{G^0G^0}(k^2_1) + \widehat{\Pi}^{G^0G^0}(k^2_2)\,
\Big]\, ,\nonumber
\end{eqnarray}
where $\widehat{\Gamma}^{HZG^0}_\mu$  and $\widehat{\Gamma}^{HG^0G^0}$
are  the one-loop PT $HZG^0$ and  $HG^0G^0$ vertices, respectively. In
this derivation,  one should also make use  of the PT WI involving the
$ZG^0$- and  $G^0G^0$- self-energies:  $\widehat{\Pi}^{ZG^0}_\mu (k) =
-iM_Zk_\mu\, \widehat{\Pi}^{G^0G^0}(k^2)/k^2$.

In  conclusion,  we have  explicitly demonstrated that   within the PT
resummation approach, any resonant Higgs-mediated amplitude contains a
gauge-independent universal part,     which is invariant   under   the
renormalization   group  and satisfies   the  optical and  equivalence
theorems individually.  It would be of great phenomenological interest
to confront  the theoretical predictions   for this universal quantity
against data obtained from future Higgs-boson experiments.
 
\samepage

\newpage 

%******************************************************************
%%%Figure 1
%******************************************************************

%******************************************************************
%%%Figure 2
%******************************************************************

\end{document}